\documentclass[11pt]{article}
\pdfoutput=1
\usepackage{fullpage}

\usepackage{amsmath,amssymb}
\usepackage{hyperref}
\usepackage{graphicx}

\usepackage[utf8]{inputenc}
\DeclareUnicodeCharacter{00A0}{~}

\begin{document}

\renewcommand{\topfraction}{1} 
\renewcommand{\bottomfraction}{1}
\renewcommand{\floatpagefraction}{1}
\renewcommand{\textfraction}{0}


\newcommand{\ie}{i.e.,\ }
\newcommand{\eg}{e.g.,\ }

\newcommand{\const}{\operatorname{const.}} 


\newcommand{\rmd}{\,\mathrm{d}}

\newcommand{\Tr}{\operatorname{tr}}

\newcommand{\e}[1]{\operatorname{e}^{#1}}

\newcommand{\op}{\mathcal{O}}

\newcommand{\vev}[1]{\left\langle #1 \right\rangle}
\newcommand{\fvev}[1]{\langle #1 \rangle}
\newcommand{\jvev}[1]{\left\langle j\left| #1 \right| j \right\rangle}

\newcommand{\comb}[2]{\begin{pmatrix} #1\\#2\end{pmatrix}}

\newcommand{\Lag}{\mathcal{L}}
\newcommand{\Ham}{\mathcal{H}}

\newcommand{\Order}{\mathcal{O}}

\newcommand{\Mpl}{M_{\text{P}}}
\newcommand{\Lpl}{L_{\text{P}}}

\newcommand{\pois}[2]{\left\{#1,#2\right\}}
\newcommand{\dirac}[2]{\pois{#1}{#2}_D}
\newcommand{\commut}[2]{\left[#1,#2\right]}

\newcommand{\be}{\mathbf{e}}
\newcommand{\bx}{\mathbf{x}}
\newcommand{\by}{\mathbf{y}}
\newcommand{\bv}{\mathbf{v}}
\newcommand{\bw}{\mathbf{w}}
\newcommand{\bk}{\mathbf{k}}
\newcommand{\br}{\mathbf{r}}
\newcommand{\bj}{\mathbf{j}}
\newcommand{\bn}{\mathbf{n}}
\newcommand{\bY}{\mathbf{Y}}
\newcommand{\bX}{\mathbf{X}}
\newcommand{\bV}{\mathbf{V}}
\newcommand{\bW}{\mathbf{W}}
\newcommand{\bA}{\mathbf{A}}
\newcommand{\bB}{\mathbf{B}}
\newcommand{\bE}{\mathbf{E}}
\newcommand{\bP}{\mathbf{P}}
\newcommand{\bPsi}{\mathbf{\Psi}}
\newcommand{\bPhi}{\mathbf{\Phi}}
\newcommand{\bpi}{\mathbf{\pi}}

\newcommand{\bval}[1]{\overline{#1}}

\newcommand{\Nq}{N_{(q)}}
\newcommand{\Nj}{N_{(j)}}
\newcommand{\Nper}{N_{(\perp)}}
\newcommand{\Np}{N_{(\parallel)}}

\newcommand{\htt}{\hat{h}}
\newcommand{\ptt}{\hat{\pi}}

\newcommand{\re}{\operatorname{Re}}
\newcommand{\im}{\operatorname{Im}}

\newcommand{\LegendreP}{\operatorname{P}}
\newcommand{\BesselJ}{\operatorname{J}}
\newcommand{\BesselH}{\operatorname{H}}
\newcommand{\BesselN}{\operatorname{N}}
\newcommand{\BesselI}{\operatorname{I}}
\newcommand{\BesselK}{\operatorname{K}}

\newcommand{\Vol}{\mathcal{V}}

\begin{center}

{\Large \textbf{Photons in a Ball}}\\[2em]

\renewcommand{\thefootnote}{\fnsymbol{footnote}}
Wolfgang M{\"u}ck${}^{a,b}$\footnote[1]{wolfgang.mueck@na.infn.it}\\[2em]
\renewcommand{\thefootnote}{\arabic{footnote}}
${}^a$\emph{Dipartimento di Fisica ``Ettore Pancini'', Universit\`a degli Studi di Napoli "Federico II"\\ Via Cintia, 80126 Napoli, Italy}\\[1em] 
${}^b$\emph{Istituto Nazionale di Fisica Nucleare, Sezione di Napoli\\ Via Cintia, 80126 Napoli, Italy}\\[2em]

\abstract{The electromagnetic field inside a spherical cavity of large radius $R$ is considered in the presence of stationary charge and current densities. $R$ provides infra-red regularization while maintaining gauge invariance. The quantum ground state of physical photons forming the magnetic field is found to be a coherent state with a definite mean occupation number. The electric field, which is determined by the Gauss law constraint, is maintained by a minimum uncertainty coherent state, according to the projection operator approach to the quantization of constrained systems. The mean occupation number of this state is proportional to the square of the total charge. The results confirm formulae obtained previously from a calculation with a finite photon mass for infra-red regularization.}

\end{center}


\section{Introduction and summary}

The quantum $N$-portrait of black holes and the corpuscular nature of gravity recently developed by Dvali, Gomez and collaborators \cite{Dvali:2010jz, Dvali:2011aa, Dvali:2012rt, Dvali:2012en, Dvali:2013vxa, Dvali:2013eja, Dvali:2015cwa, Dvali:2015aja, Dvali:2015rea} are interesting approaches to tackling profound questions in quantum gravity, from ultra-violet finiteness over black hole entropy to the information paradox. They also may help to shed light on dark energy: Applying the same ideas to the observable universe gives an estimate of the dark energy density which is very close to the observed value \cite{Binetruy:2012kx}. Related developments can be found in \cite{Vikman:2012bx, Kovner:2012yi, Alberte:2012is, Flassig:2012re, Berkhahn:2013woa, Casadio:2013hja, Muck:2014kea, Casadio:2014vja, Casadio:2015xva, Casadio:2015bna}.

A pressing question in this context is how to identify, from an underlying microscopic quantum theory, the $N$ quantum constituents of a given semiclassical gravitational configuration. Evidence from string theory and supergravity shows that the degeneracy of soft graviton states of order $e^N$ implied by the quantum $N$-portrait is needed in order for the S-matrix of graviton-graviton scattering to be unitary \cite{Dvali:2014ila}. More recently, a proposal for how to identify the quantum constituents of solitons has been made \cite{Dvali:2015jxa}, but it is too early to draw conclusions from this for gravity. 

An obvious approach to the above question is to view classical fields as coherent quantum states and to identify $N$ with the mean occupation number. This approach is based upon an intuitive understanding of quantum field theory as an infinite collection of harmonic oscillators, for which the concept of coherent states is most straightforward. However, in this respect, gravity poses two challenges. First, gravity is a constrained dynamical system as a consequence of gauge-invariance. Second, the long-range nature of gravity renders classical field configurations not square-integrable and, therefore, unsuitable as physical quantum states.\footnote{This is similar to the situation of the topological solitons in \cite{Dvali:2015jxa}. Therefore, following the approach of that paper, one may ponder about a possible convolution of a topological sector and an energy sector in gravity.} 

Both challenges are shared by electrodynamics, and it has been argued before that understanding the analogous question---how many photons are bound by electric charges and currents---would yield answers that could be generalized to gravity \cite{Barnich:2010bu, Mueck:2013mha, Mueck:2013wba}. In \cite{Mueck:2013wba}, this question has been answered circumventing the above problems by adding a small photon mass and a Coulomb gauge fixing term to the Maxwell Lagrangian. The former acts as an infra-red regulator, breaks gauge-invariance and makes the constraints second-class, while the latter ensures that the dispersion relation for longitudinal photons differs from that of the physical transverse photons. In the limit where the photon mass is removed, the spectrum of the longitudinal modes is pushed to infinity, so that they ``freeze''. Nevertheless, the classical electrostatic field around a given charge distribution really is represented by a coherent state of longitudinal photons, and the mean occupation number of that state is
\begin{equation}
\label{intro:nq}
	N_q = c \frac{q^2}{4\pi\hbar}~,
\end{equation}
where $q$ is the total charge and $c$ is a numerical constant that depends on the coefficient of the gauge fixing term. This fact makes the result \eqref{intro:nq} rather suspicious and indicates that $N_q$, although calculable, is not a physical observable. In contrast, the number of transverse, \ie physical photons in the static magnetic field generated by a stationary current density $\bj(\bx)$ was found to be
\begin{equation}
\label{intro:nj}
	N_j = \frac1{(2\pi)^2 \hbar} \int \rmd^3 x\, \rmd^3 y\, j^i(\bx) j^j(\by) 
	\frac{(x-y)_i(x-y)_j}{|\bx-\by|^2}~.
\end{equation}
This result is independent of the gauge parameter, but one may still wonder whether it is meaningful or just an artifact of the infra-red regularization via a photon mass. 

It is, therefore, desirable to have a second, independent check of the above equations \eqref{intro:nq} and \eqref{intro:nj}, and this is what we will pursue in this paper. Most importantly, we want to get rid of the gauge-invariance breaking photon mass. For infra-red regularization, we consider the system in a finite volume, namely a ball of radius $R$, and take $R\to \infty$ at the end of the day. Boundary conditions at $r=R$ are fixed by assuming no influence of the outside region on the fields inside the ball.

The outline of the paper is as follows. Our starting point is the Maxwell action
\begin{equation}
\label{intro:action}
	S = \int \rmd t \int\limits_{r\leq R} \rmd^3 x \left(\frac12 \bE^2 -\frac12 \bB^2 + j^\mu A_\mu\right)~,
\end{equation}
where $A_\mu=(-\Phi,\bA)$ denotes the electromagnetic vector potential, and $\bE=-\nabla \Phi -\partial_t \bA$ and $\bB= \nabla \times \bA$ are the usual electric and magnetic field strengths, respectively. The four-current $j^\mu$ is assumed to be stationary, in addition to being conserved, so that the Hamiltonian is explicitly time-independent. The action \eqref{intro:action} is manifestly gauge invariant, provided the current density across the boundary is zero. In Sec.~\ref{quant}, the system is analized in the Hamiltonian formalism and quantized.   Inside the ball, all fields can be expanded in terms of a discrete basis of eigenfunctions of the Laplace operator. Using such a basis, it is quite evident which components can be explicitly set to zero by a gauge transformation. The resulting system contains two standard harmonic oscillators describing the two physical photons, coupled to the stationary current density, while the dynamics of the field $\Phi$ is fully constrained as expected. What is important is that, because of the explicit gauge fixing, the constraints are second-class. In the projection operator approach to the quantization of constrained systems  \cite{Klauder:1996nx, Govaerts:1996zn}, a totally constrained canonical pair of variables is physically implemented by the coherent, minimum uncertainty state associated with the classical values of the constrained variables.\footnote{Such a state is not an eigenstate of the Hamiltonian. Rather, the time evolution operator involves a projection onto this state at every moment. This is the quantum equivalent of the fact that, in classical dynamics, the Lagrange multipliers for second-class constraints are dynamically determined functions of time, ensuring that the system remains on the constraint subspace. In contrast, first-class constraints can be imposed as initial conditions \cite{Klauder:1996nx}.} 
We identify the number of $\Phi$-quanta with the mean occupation number of this coherent state. The numerical constant $c$ in \eqref{intro:nq} is seen to arise from the arbitrariness of choosing a length scale when transforming the phase space variables to an oscillator basis. In the physical sector, we find that the ground state is a coherent state of photons which describes the classical magnetic field generated by the stationary electric currents. We shall verify in Sec.~\ref{loop}, using a rather technical calculation, that the mean occupation number associated with such a ground state agrees with \eqref{intro:nj} in the case of a simple configuration of current densities. Moreover, we provide a check of \eqref{intro:nq}. Technical details of the mode decomposition and a discussion of the treatment of boundary values are deferred to the Appendix.

The interpretation of $N_q$ that emerges from our results is that it is a measure (albeit an ambiguous one) of the number of quanta needed to keep the dynamics on the constrained subspace. For gravity, where $N=(M/\Mpl)^2$, the question that must be answered is how such quanta become accessible to the S-matrix, as evidenced in \cite{Dvali:2014ila} and through black hole formation.

To end, let us briefly outline how the projection operator approach may be used to derive \eqref{intro:nq} directly when quantizing in Minkowski space. This points to a connection with the recently discussed soft photon theorem as the Ward identity of an asymptotic symmetry of QED \cite{Kapec:2015ena, He:2014cra} and, in the case of gravity, to a relation with gravitational memory and the BMS symmetry group \cite{Strominger:2014pwa}, see also \cite{Dvali:2015rea}. Consider the asymptotic quantization of the electromagnetic field in retarded radial gauge. Using the notation of \cite{Kapec:2015ena}, the free boundary data at leading order is $F_{ru}^{(2)}=-A_u^{(1)}$, the constant part of which (on $S^2$) is proportional to the total electric charge in the system. However, $F_{ru}^{(2)}$ is the canonical conjugate of a component of $A_r$, which has been gauge fixed to zero. Hence, the quantum time evolution involves a projection onto the minimum uncertainty coherent state associated with the classical values. The mean occupation number of this state is trivially given by \eqref{intro:nq}.

\section{Electrodynamics inside a ball}
\label{quant}

In this section, we analyze the action \eqref{intro:action} in the Hamiltonian formalism, gauge fix, quantize. To do so, we expand all fields in a complete and orthonormal basis of eigenfunctions of the Laplace operator. For details of the expansion, including a discussion of subtle issues involving boundary values, we refer to the Appendix. The choice of boundary conditions is a crucial aspect of the field dynamics. Boundary conditions may be implemented by zero modes, which may or may not be independent of the other modes, and must be such that the on-shell variation of the action,
\begin{equation}
\label{quant:action.var}
	\delta S = \int \rmd t \int\limits_{r=R} \rmd^2 x 
	\left[ -\bn \cdot \bE\, \delta \Phi + (\bn \times \bB) \cdot \delta \bA \right]~,
\end{equation}
vanishes. Here, $\bn=\br/r$ denotes the spacial unit normal vector on the boundary. Using the notation introduced in the Appendix, one easily realizes that $\delta S=0$ is satisfied for the simplest choice of a complete basis, in which $\{k\}$ is chosen as \eqref{appexp:Bessel.bc} and  
\begin{equation}
\label{quant:no.modes}
	\bval{\Phi}_{lm}= \bval{A}^V_{lm} = \bval{A}^X_{lm} =0~.
\end{equation}
These are precisely the coefficients of the zero modes which are not independent. That these modes are not relevant for our purposes can also be seen from the fact that they do not appear in the source term $(j^\mu A_\mu)$ and, therefore, would be physically determined by the field dynamics in the region outside the ball and, potentially, by matching conditions at the boundary. We assume these to be trivial. 

Similarly, we expand the charge and current densities, $\rho=j^0$ and $\bj$, respectively, in terms of the independent modes. Charge conservation, stationarity and the absence of currents crossing the boundary (needed for gauge invariance) imply 
\begin{equation}
\label{quant:no.j}
	\widetilde{j}_{lmk}= \bval{j}^W_{lm}=0~.
\end{equation}
Let us also gauge fix. As is evident from \eqref{appexp:grad}, we can use a gauge transformation to set\footnote{Setting to zero $\bval{A}^W_{lm}$ is subtle in connection with the boundary condition for $\bval{\Phi}_{lm}$, because this eliminates the independent zero modes of the electric field. One may argue again that these components are not sourced and, therefore, would be determined by the physics outside the ball.}
\begin{equation}
\label{quant:gaugefix}
	\widetilde{A}_{lmk}= \bval{A}^W_{lm}=0~.
\end{equation}

With these preliminaries, the Lagrangian takes the following simple form,\footnote{That these expressions and others that follow are real stems from the reality condition \eqref{appexp:reality}.}
\begin{subequations}
\begin{align}
\label{em:lag}
	L &= L^X +\widehat{L} +L^\Phi~,\\
	L^X &= \sum_{lmk} \left[ \frac12 \left|\partial_t A^X_{lmk}\right|^2 -\frac12 k^2 \left| A^X_{lmk} \right|^2 
	+ (j^X_{lmk})^\ast A^X_{lmk} \right]~,\\
	\widehat{L} &= \sum_{lmk} \left[ \frac12 \left|\partial_t \widehat{A}_{lmk}\right|^2 -\frac12 k^2 \left| \widehat{A}_{lmk} \right|^2 
	+ (\widehat{j}_{lmk})^\ast \widehat{A}_{lmk} \right]~,\\
	L^\Phi &= \sum_{lmk} \left[ \frac12 k^2 \left|\Phi_{lmk} \right|^2 - (\rho_{lmk})^\ast \Phi_{lmk} \right]~.
\end{align}
\end{subequations}
The standard treatment gives rise to the Hamiltonian\footnote{For notational simplicity, we define the canonical momenta by $\pi = \frac{\delta L}{\delta(\partial_t \phi^\ast)}$.}
\begin{subequations}
\begin{align}
\label{em:ham}
	H &= H^X +\widehat{H} +H^\Phi~,\\
\label{em:ham.X}
	H^X &= \sum_{lmk} \left[ \frac12 \left|\pi^X_{lmk}\right|^2 +\frac12 k^2 \left| A^X_{lmk} \right|^2 
	- (j^X_{lmk})^\ast A^X_{lmk} \right]~,\\
	\widehat{H} &= \sum_{lmk} \left[ \frac12 \left|\widehat{\pi}_{lmk}\right|^2 +\frac12 k^2 \left| \widehat{A}_{lmk} \right|^2 
	- (\widehat{j}_{lmk})^\ast \widehat{A}_{lmk} \right]~,\\
\label{em:ham.Phi}	
	H^\Phi &= - L^\Phi = \sum_{lmk} \left[ -\frac12 k^2 \left|\Phi_{lmk} \right|^2 + (\rho_{lmk})^\ast \Phi_{lmk} \right]~,
\end{align}
\end{subequations}
with the non-zero Poisson brackets
\begin{equation}
\label{em:pois}
	\big\{A^X_{lmk},(\pi^X_{l'm'k'})^\ast\big\} = 
	\big\{\widehat{A}_{lmk},(\widehat{\pi}_{l'm'k'})^\ast\big\} = 
	\big\{\Phi_{lmk},(\pi_{l'm'k'})^\ast\big\} = \delta_{ll'}\delta_{mm'}\delta_{kk'}~.	
\end{equation}

Finding the ground state for $H^X$ and $\widehat{H}$, which describe canonical harmonic oscillators, is straightforward. Consider $H^X$. After quantization, the Hamiltonian \eqref{em:ham.X} is diagonalized by the transformation\footnote{We omit any notation on the ladder operators that would distinguish between the different field components. The somewhat unusual signs in \eqref{em:A.X} arise from the reality condition \eqref{appexp:reality}.}
\begin{equation}
\label{em:A.X}
	A^X_{lmk} = \sqrt{\frac{\hbar}{2k}} \left[ a_{lmk} + (-1)^m a_{l(-m)k}^\dagger \right]~, \quad
	\pi^X_{lmk} = -i \sqrt{\frac{\hbar k}{2}} \left[ a_{lmk} - (-1)^m a_{l(-m)k}^\dagger \right]~,
\end{equation}  
with the ladder operators satisfying
\begin{equation}
\label{em:a.comm}
	\commut{a_{lmk}}{a_{l'm'k'}^\dagger} = \delta_{ll'}\delta_{mm'}\delta_{kk'}~.
\end{equation}
The Hamiltonian \eqref{em:ham.X} becomes, normal ordered and with the zero point energy dropped,
\begin{equation}
\label{em:ham.Phi.n}
	H^X = \sum_{lmk} \left[ \hbar k 
	\left( a_{lmk}^\dagger -\frac1{\sqrt{2\hbar k^3}} (j^X_{lmk})^\ast\right) 
	\left( a_{lmk} -\frac1{\sqrt{2\hbar k^3}} j^X_{lmk}\right) 
	- \frac1{2k^2} \left|j^X_{lmk}\right|^2 \right]~.
\end{equation}
Therefore, the ground state of $H^X$ in the presence of a non-zero source is a coherent state of photons,
\begin{equation}
\label{em:coh.X}
	a_{lmk} |j\rangle = \frac{j^X_{lmk}}{\sqrt{2\hbar k^3}} |j\rangle~,
\end{equation}
with the mean occupation number given by
\begin{equation}
\label{em:N.X}
	N^X = \sum_{lmk} \frac1{2\hbar k^3} \left| j^X_{lmk} \right|^2~.
\end{equation}
The analysis for $\widehat{H}$ is identical.

The dynamics of the electric field is fully determined by the constraints
\begin{equation}
\label{em:constraints}
	\pi_{lmk} \approx 0~,\qquad \big\{ H^\Phi,\pi_{lmk} \big\} = -k^2 \Phi_{lmk} + \rho_{lmk} \approx 0~.
\end{equation}
Because the constraints are second-class, they cannot be imposed as operator identities on the full Hilbert space. Whereas Dirac's formalism would effectively discard this canonical pair, the projection operator approach to the quantization of constrained systems \cite{Klauder:1996nx, Govaerts:1996zn} treats the constraints as classical equations, which must be approximated as precisely as possible  by a quantum state. For second-class constraints, this means a projection onto a minimun uncertainty coherent state.
Introduce ladder operators satisfying \eqref{em:a.comm} by 
\begin{equation}
\label{em:Phi.pi}
	\Phi_{lmk} = \sqrt{\frac{\hbar L_{lmk}}{2}} \left[ a_{lmk} + (-1)^m a_{l(-m)k}^\dagger \right]~,\quad
	\pi_{lmk} = -i \sqrt{\frac{\hbar}{2L_{lmk}}} \left[ a_{lmk} - (-1)^m a_{l(-m)k}^\dagger \right]~,
\end{equation}  
where $L_{lmk}$ are arbitrary, but of unit length. This is different from the dynamical degrees of freedom \eqref{em:A.X}, where this factor was uniquely determined by the diagonalization of the Hamiltonian. Here, it is natural to construct $L_{lmk}$ as a combination of $R$ and $1/k$, with non-negative coefficients of order unity that may depend on $l$ and $m$,
\begin{equation}
\label{em:L}
	L_{lmk} = \alpha_{lm} \frac1k +\beta_{lm} R~.
\end{equation}
Then, using \eqref{em:Phi.pi} and \eqref{em:constraints}, the mean occupation number of the coherent state implementing the constraints is found to be
\begin{equation}
\label{em:N.Phi}
	N^\Phi = \sum_{lmk} \frac{\left|\rho_{lmk}\right|^2}{2\hbar(\alpha_{lm}k^3 +\beta_{lm} R k^4)}~. 
\end{equation}

In the next section, we will check with simple examples that \eqref{em:N.X} and \eqref{em:N.Phi} agree with \eqref{intro:nj} and \eqref{intro:nq}, respectively, when $R\to \infty$. 
 
\section{Checks of photon number formulae}
\label{loop}

Here we check that our results \eqref{em:N.X} and \eqref{em:N.Phi} agree, for simple configurations of current and charge densities and in the $R\to \infty$ limit, with the general formulae \eqref{intro:nj} and \eqref{intro:nq}, respectively.

\subsection{Magnetic field}
Consider a current $I$ running in a loop of radius $\rho$. In spherical coordinates, the current density can be given by 
\begin{equation}
\label{loop:j.loop}
	\bj(\br) =  \frac{I}r \delta(\theta-\pi/2)\, \delta(r-\rho)\, \be_\phi~.
\end{equation}
The general formula $\eqref{intro:nj}$ readily yields 
\begin{equation}
\label{loop:Nj.loop}
	N_j = \frac{(I\rho)^2}{2 \hbar}~.
\end{equation} 

We start by decomposing the current density \eqref{loop:j.loop}. One easily realizes that the only non-zero components are\footnote{While   $\bY_{lm}$ is orthogonal to $\be_\phi$, the $\be_\phi$ component of $\bPsi_{lm}$ vanishes upon the $\phi$-integration.}
\begin{align}
\notag
	j^{X}_{lmk} &= 
	  \frac1{c_{lk}\sqrt{l(l+1)}} \int \rmd^3r\, j_l(kr)\, \bj(\br) \cdot \left(\br \times \nabla Y^\ast_{lm} \right)\\
\notag
	  &= \frac1{c_{lk}\sqrt{l(l+1)}} \int \rmd^3r\, j_l(kr) \nabla Y^\ast_{lm} \cdot\left[ \bj(\br) \times \br \right]\\
\label{loop:j3.1}
	  &= -\frac{I}{c_{lk}\sqrt{l(l+1)}} \rho j_l(k\rho) \int\limits_0^{2\pi} \rmd \varphi 
	  \left. \partial_\theta Y_{lm}^\ast \right|_{\theta=\pi/2}~.
\end{align}
With the explicit expression of the spherical harmonics
\begin{equation}
\label{loop:Y.expl}
	Y_{lm}(\Omega) = (-1)^m \sqrt{\frac{(2l+1)(l-m)!}{4\pi (l+m)!}} \LegendreP_l^m(\cos\theta) \e{im\varphi}~,
\end{equation}
where $\LegendreP^m_l(x)$ denote the associated Legendre polynomials, \eqref{loop:j3.1} becomes
\begin{align}
\notag
	j^X_{lmk} &= \frac{I}{c_{lk}}  \delta_{m0} \sqrt{\frac{\pi(2l+1)}{l(l+1)}}\, \rho j_l(k\rho) \LegendreP_l'(0)\\
\label{loop:j3.2}
	&= \frac{I}{c_{lk}} \delta_{m0} \sqrt{\frac{\pi(2l+1)}{l(l+1)}}\, \rho j_l(k\rho) \begin{cases}
	(-1)^n \frac{(2n+1)!!}{2^n n!} \qquad & \text{for $l=2n+1$,}\\
	0 & \text{for $l=2n$.} \end{cases}
\end{align}

Hence, the mean photon number in the magnetic field \eqref{em:N.X} is 
\begin{align}
\notag
	N^X &= \sum_{lmk} \frac1{2\hbar k^3} \left| j^X_{lmk} \right|^2 \\
\label{loop:N.loop}
	&= \frac{(I\rho)^2}{2\hbar} \sum_{n=0}^\infty \frac{2\pi(4n+3)}{(2n+1)(2n+2)} \left[ \frac{(2n+1)!!}{2^n n!}\right]^2
		\sum_{x_n} \frac1{x_n^3} \left[ \frac{j_{2n+1}(x_n\rho/R)}{j_{2n+2}(x_n)} \right]^2~,
\end{align}
where the $x_n$ denote the positive zeros of $j_{2n+1}(x)$. We have used also \eqref{appexp:clk}. To verify \eqref{loop:Nj.loop}, we have to show that 
\begin{equation}
\label{loop:sum.to.verify}
	\sum_{n=0}^\infty \frac{2\pi(4n+3)}{(2n+1)(2n+2)} \left[ \frac{(2n+1)!!}{2^n n!}\right]^2
		\lim_{R\to\infty} \sum_{x_n} \frac1{x_n^3} \left[ \frac{j_{2n+1}(x_n\rho/R)}{j_{2n+2}(x_n)} \right]^2 = 1~.
\end{equation}
Let us consider the sum over $x_n$. Expressing the spherical Bessel functions in terms of Bessel functions, we have
\begin{equation}
\label{loop:sum1}
	\sum_{x_n} \frac1{x_n^3} \left[ \frac{j_{2n+1}(a x_n)}{j_{2n+2}(x_n)} \right]^2 = \frac1a 
	\sum_{x_n} \frac1{x_n^3} \left[ \frac{J_{2n+3/2}(a x_n)}{J_{2n+5/2}(x_n)} \right]^2~,
\end{equation}
and we can identify the $x_n$ also as the positive zeros of $J_{2n+3/2}(x)$. The sum on the right hand side of \eqref{loop:sum1} is very similar to the one appearing in the Kneser-Sommerfeld formula in Watson's treatise on Bessel functions \cite{Watson}. This formula, however, is known to be incorrect \cite{Hayashi:1981, Vaccaro:1998}, but we can apply the method used (wrongly) in the proof of the formula to obtain   our result.\footnote{Compared to our analysis, the proof in the book wrongly states that the contributions to the contour integral from the upper and lower imaginary axes cancel each other, which is not the case.} The method goes back to \cite{Kneser, Sommerfeld}. 
\begin{figure}
\begin{center}
	\includegraphics[width=0.2\textwidth]{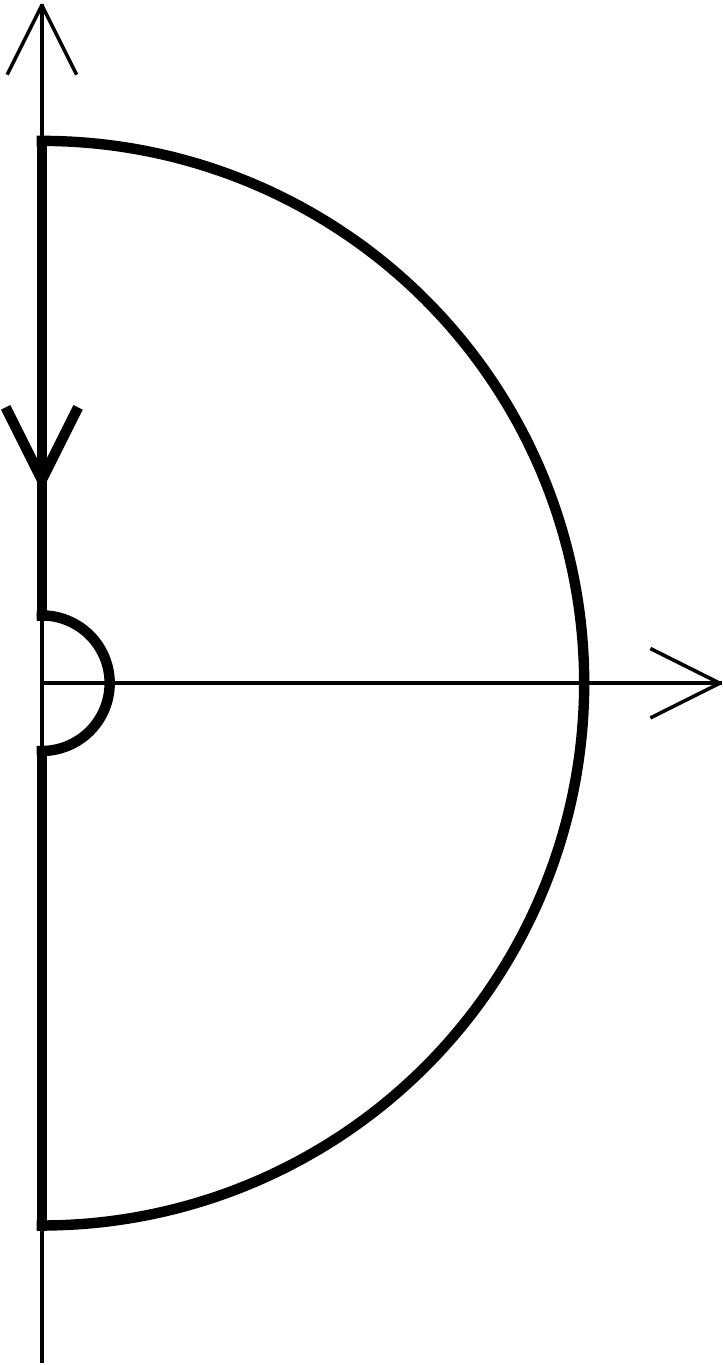}
\end{center}
\caption{Contour in the complex plane used to evaluate the sum \eqref{loop:sum1}. The radii of the large and small semicircles are intended as the limits to $\infty$ and $0$, respectively. \label{loop:contour}}
\end{figure}

For $\nu>1/2$, which holds in our case, consider the following function of a complex variable $z$,\footnote{$\BesselJ_\nu(z)$, $\BesselN_\nu(z)$, $\BesselH^{(1)}_\nu(z)$ and $\BesselH^{(2)}_\nu(z)$ denote the Bessel functions of the first, second and third kind, respectively.}
\begin{equation}
\label{loop:Fz}
	F(z) = \frac{\BesselH_\nu^{(1)}(az)\BesselH_{\nu}^{(2)}(z)-\BesselH_\nu^{(2)}(az)\BesselH_{\nu}^{(1)}(z)}{z^2} 
	\frac{\BesselJ_\nu(az)}{\BesselJ_{\nu}(z)}~.
\end{equation}
Calculate the integral of $F(z)$ on the loop contour depicted in Fig.~\ref{loop:contour}. The residue theorem yields 
\begin{equation}
\label{loop:cont.int}
	\oint \rmd z\, F(z) = -8 \sum_{x_n} \frac1{x_n^3} \left[ \frac{J_{\nu}(a x_n)}{J_{\nu+1}(x_n)} \right]^2~,
\end{equation}
where the sum on the right hand side is over all the positive zeros of $J_{\nu}(x)$. On the right hand side, we recognize the expression we need in \eqref{loop:sum1}. The contour integral on the left hand side must be evaluated directly. It is easy to see that the integral over the large semicircle vanishes in the infinite radius limit. In contrast to the derivation of the (corrected) Kneser-Sommerfeld formula \cite{Hayashi:1981, Vaccaro:1998}, the contributions from the upper and lower imaginary axes does not cancel, but are equal to each other. In addition, the contribution from the small semicircle, which is divergent in the zero radius limit, is easily found by considering the asymptotic behaviour of the integrand. Thus, after some manipulation, one finds
\begin{equation}
\label{loop:Fint}
	\oint \rmd z\, F(z) = \lim_{\epsilon \to 0} \Bigg\{ \frac8\pi \int\limits_\epsilon^\infty \frac{\rmd y}{y^2} 
	\left[ \BesselI_\nu(ay) \BesselK_\nu(ay)-\frac{\BesselI_\nu(ay)^2\BesselK_{\nu}(y)}{\BesselI_{\nu}(y)} \right]
	-\frac4{\epsilon \nu \pi}\left(1 -a^{2\nu}\right) \Bigg\}
\end{equation} 
The last term is the contribution from the small semicircle contour. To continue, we integrate by parts after writing $y^{-2}=-\partial_y y^{-1}$ and find that the boundary term cancels the term from the semicircle contour. The remaining integral is finite in the $\epsilon \to 0$ limit,
\begin{equation}
\label{loop:Fint1}
	\oint \rmd z\, F(z) = \frac8\pi \int\limits_0^\infty \frac{\rmd y}{y} \left\{
	a \left[\BesselK_\nu(ay) \BesselI_{\nu+1}(ay) - \BesselK_{\nu-1}(ay) \BesselI_\nu(ay) \right] 
	- \partial_y \frac{\BesselI_\nu(ay)^2\BesselK_{\nu}(y)}{\BesselI_{\nu}(y)} \right\} 
\end{equation} 
We are interested in the $a\to 0$ behaviour. Rescaling the integration variable by $y\to y/a$ shows that the last term in the integral is exponentially suppressed in this limit, while the other two terms can be readily integrated, resulting in
\begin{equation}
\label{loop:Fint2}
	\oint \rmd z\, F(z) = -\frac{4a}{\pi} \left(\nu^2-\frac14\right)^{-1} +\cdots~,
\end{equation}	
where the ellipses indicate the exponentially suppressed terms. 

Then, combining \eqref{loop:Fint2}, \eqref{loop:cont.int} and \eqref{loop:sum1} yields (remember $\nu=2n+\frac32$)
\begin{equation}
\label{loop:sum2}
  \lim\limits_{a\to 0}\sum_{x_n} \frac1{x_n^3} \left[ \frac{j_{2n+1}(a x_n)}{j_{2n+2}(x_n)} \right]^2 = \frac1{2\pi(2n+1)(2n+2)}~,
\end{equation}
and the left hand side of \eqref{loop:sum.to.verify} becomes
\[ \sum_{n=0}^\infty \frac{(4n+3)}{(2n+2)^2} \left[ \frac{(2n-1)!!}{2^n n!}\right]^2 = 
   \sum_{n=0}^\infty \frac{(4n+3)}{(2n+2)^2} \left| \LegendreP_{2n}(0) \right|^2~.
\]
Finally, using a recursion relation for Legendre polynomials, this can be rewritten as
\[ \sum_{n=0}^\infty \left[ 1- \frac{(2n+1)^2}{(2n+2)^2} \right] \left| \LegendreP_{2n}(0) \right|^2 = 
   \sum_{n=0}^\infty \left[ \left| \LegendreP_{2n}(0) \right|^2 - \left| \LegendreP_{2n+2}(0) \right|^2 \right] 
   = \left| \LegendreP_0(0) \right|^2 = 1~,  \]
which proves \eqref{loop:sum.to.verify}.

\subsection{Electric field}
Consider a spherical shell of radius $\rho$ and total charge $Q$, described by the charge density
\begin{equation}
\label{el:charge.density}
	\rho = \frac{Q}{4\pi r^2} \delta(r-\rho)~.
\end{equation}
Expanding \eqref{el:charge.density} yields the non-zero mode coefficients
\begin{equation}
\label{el:rho.k}
	\rho_{00k} = \frac{Q}{\sqrt{2\pi R^3}}\frac{j_0(k\rho)}{j_1(kR)}~,
\end{equation}
so that the mean number of quanta \eqref{em:N.Phi} becomes
\begin{equation}
\label{el:N.Phi}
	N^\Phi = \frac{Q^2}{4\pi \hbar} \sum_{x_n} \frac1{\alpha x_n^3 +\beta x_n^4} \left[\frac{j_0(x_n\rho/R)}{j_1(x_n)}\right]^2~, 
\end{equation}
with $x_n=n\pi$ ($n=1,2,\ldots$) being the positive zeros of $j_0(x)$. 

If we show that the sum in \eqref{el:N.Phi} is finite, we would have verified \eqref{intro:nq} for this simple case. It is evident that the constant $c$ in \eqref{intro:nq} arises from the freedom to choose $\alpha$ and $\beta$. First, let us show that $\beta$ must be non-zero in order for $c$ to be finite. For $\beta=0$, the sum is of the same form as we discussed in the previous subsection, but here we have $\nu=1/2$, so that the result \eqref{loop:Fint2} is divergent. We mention that, in the case of regularization with a finite photon mass \cite{Mueck:2013wba}, $c$ would have diverged if the gauge fixing term had been removed, \ie for a Proca Lagrangian. 

The sum can be evaluated exactly for the special case $\alpha=0$. In that case, after inserting the explicit expressions for the spherical Bessel functions, we obtain\footnote{The sums can be found in \cite{Gradshteyn}.}
\begin{equation}
\label{el:c.alpha.0}
	c= \frac{R^2}{2\pi^4 \beta \rho^2} \sum_{n=1}^\infty \frac{1-\cos(2\pi n \rho/R)}{n^4} = \frac1{6\beta} \left(1-\frac{\rho}{R}\right)^2~.
\end{equation}
This vanishes for $r=R$ as expected from \eqref{el:N.Phi}, \ie when the charge is removed from the ball. The limit $R\to0$ is trivial.

In the general case, we have
\begin{equation}
\label{el:c.generic}
	c= \frac{R^2}{\beta \pi^4 \rho^2} \sum_{n=1}^\infty \frac{\sin^2(\pi n\rho/R)}{n^4+\frac{\alpha}{\pi \beta} n^3}~.
\end{equation}
Taking the $R\to \infty$ limit first yields
\begin{equation}
\label{el:c.generic.limit}
	c= \frac{1}{\beta \pi^2} \sum_{n=1}^\infty \frac{1}{n\left(n+\frac{\alpha}{\pi \beta}\right)} 
	= \frac1{\pi \alpha} \left[\psi\left(1+\frac{\alpha}{\pi \beta} \right) -\psi(1) \right] 
	= \frac1{\pi \alpha} \sum_{k=2}^\infty (-1)^k \zeta(k) \left(\frac{\alpha}{\pi \beta} \right)^{k-1}~,
\end{equation}
under the condition $\frac{\alpha}{\beta \pi}>-1$. The limiting cases $\beta=0$ and $\alpha=0$ are easily reproduced.

Finally, let us argue that, for a given charge distribution within a finite radius, only the $l=0$ part of the sum contributes to $N^\Phi$ when the limit $R\to \infty$ is taken. Consider, for example, the $l=1$ terms. The contribution of these terms will be proportional to $\bP^2$, where $\bP$ is the total electric dipole moment, which is assumed to be finite. Hence, for dimensional reasons, the contribution of the $l=1$ terms to $N^\Phi$ will be $c_1\frac{\bP^2}{\hbar R^2}$, with a generically finite numerical constant $c_1$. Therefore, these contributions, as well as any contribution from $l>0$, vanish in the limit $R\to \infty$.

\begin{appendix}
\section{Field decomposition}
\label{app:expansion}

In this appendix, we summarize the expansion of scalar and vector fields inside a ball of radius $R$ in terms of solutions of the Helmholtz equation 
\begin{equation}
\label{appexp:helmholtz}
	\left(\nabla^2 +k^2\right) \Phi=0~, \qquad \left(\nabla^2 +k^2\right) \bA=0~.
\end{equation}

In spherical coordinates, the angular dependence of scalar and vector fields is expressed most conveniently in terms of the scalar and vector spherical harmonics, respectively. The scalar spherical harmonics $Y_{lm}(\Omega)$ are well-known functions. Vector spherical harmonics may be less familiar, but are often used in electrodynamics, for example for multipole expansions. In this paper, we adopt the definitions of the vector spherical harmonics $\bY_{lm}(\Omega)$, $\bPsi_{lm}(\Omega)$ and $\bPhi_{lm}(\Omega)$ presented in \cite{Barrera:1985}, to which we refer for details.
From these functions, we define the following mutually orthogonal and normalized combinations, similar to the conventions of \cite{Hill:1954},
\begin{subequations}
\label{appexp:vec.spher.harm}
\begin{align}
	\bV_{lm} &= \left[ (l+1)(2l+1) \right]^{-1/2} \left[-(l+1)\bY_{lm}+\bPsi_{lm}\right]~,\\
	\bW_{lm} &= \left[ l(2l+1) \right]^{-1/2} \left(l\bY_{lm}+\bPsi_{lm}\right)~,\\
	\bX_{lm} &= \left[ l(l+1) \right]^{-1/2} \bPhi_{lm}~.
\end{align}
\end{subequations}

The radial dependence of the solutions of \eqref{appexp:helmholtz} involves the spherical Bessel functions,
\begin{equation}
\label{appexp:sph.bessel}
	j_n(x) = \sqrt{\frac{\pi}{2x}} J_{n+1/2}(x)~,
\end{equation}
where $J_\nu(x)$ is a Bessel function of the first kind, and $n=0,1,2,\ldots$. We recall the standard orthogonality property \cite{Abramowitz-Stegun},
\begin{equation}
\label{appexp:Bessel.ortho}
	\int\limits_0^1 \rmd x \, x J_\nu(\alpha_m x) J_\nu(\alpha_n x) = \begin{cases}
	0&\text{for $m\neq n$,}\\
	\frac12 [J_{\nu+1}(\alpha_n)]^2 &\text{for $m=n$, $b=0$,}\\
	\frac1{2\alpha_n^2} \left(\frac{a^2}{b^2} +\alpha_n^2 -\nu^2\right)[J_\nu(\alpha_n)]^2\quad
	&\text{for $m=n$, $b\neq 0$}, \end{cases}
\end{equation}  
where the $\alpha_m$, $m=0,1,2,\ldots$ denote the positive roots of $a J_\nu(x) + bx J'_\nu(x)$, $a,b \in \mathbb{R}$, and $\nu>-1$.

Solutions of \eqref{appexp:helmholtz}, which are regular inside the ball, are 
\begin{subequations}
\label{appexp:helmholtz.sols}
\begin{align}
\label{appexp:scal.sol}
	\phi_{lmk}(\Omega,r)&= c_{lk}^{-1} Y_{lm}(\Omega) j_l(kr)  &&(l=0,1,2,\ldots),\\
\label{appexp:vec.solsa}	
	\bv_{lmk}(\Omega,r)&= c_{lk}^{-1} \bV_{lm}(\Omega) j_{l+1}(kr)  &&(l=0,1,2,\ldots),\\
\label{appexp:vec.solsb}	
	\bw_{lmk}(\Omega,r)&= c_{lk}^{-1} \bW_{lm}(\Omega) j_{l-1}(kr) &&(l=1,2,3\ldots),\\
\label{appexp:vec.solsc}	
	\bx_{lmk}(\Omega,r)&= c_{lk}^{-1} \bX_{lm}(\Omega) j_l(kr) &&(l=1,2,3\ldots),
\end{align}
\end{subequations}
where $c_{lk}$ are normalization constants, which we take to be equal with hindsight. In addition to these solutions, there are zero modes ($k=0$), for which the spherical Bessel functions $j_n(kr)$ are replaced by $r^n$,\footnote{Throughout the paper, zero-mode-related quantities are distinguished by a bar.}
\begin{subequations}
\label{appexp:helmholtz.zero.modes}
\begin{align}
\label{appexp:scal.zero}
	\bval{\phi}_{lm}(\Omega,r)&= Y_{lm}(\Omega) \left(\frac{r}{R}\right)^l  &&(l=0,1,2,\ldots),\\
\label{appexp:vec.zero.a}	
	\bval{\bv}_{lm}(\Omega,r)&= \bV_{lm}(\Omega) \left(\frac{r}{R}\right)^{l+1}  &&(l=0,1,2,\ldots),\\
\label{appexp:vec.zero.b}	
	\bval{\bw}_{lm}(\Omega,r)&= \bW_{lm}(\Omega) \left(\frac{r}{R}\right)^{l-1} &&(l=1,2,3\ldots),\\
\label{appexp:vec.zero.c}	
	\bval{\bx}_{lm}(\Omega,r)&= \bX_{lm}(\Omega) \left(\frac{r}{R}\right)^l &&(l=1,2,3\ldots).
\end{align}
\end{subequations}
Solutions involving the spherical Bessel functions $y_n(kr)$ as well as negative powers of $r$ are to be excluded by regularity. 

To form a basis, it is necessary to specify a set $\{k\}$, for which the functions \eqref{appexp:helmholtz.sols}, possibly including some of the zero modes \eqref{appexp:helmholtz.zero.modes}, are complete. We shall postpone this and first discuss properties that are independent of the choice of $\{k\}$.

Consider a generic (possibly overcomplete) expansion for real scalar fields, which reads 
\begin{equation}
\label{appexp:scalar.exp}
	\Phi(\Omega,r) = \sum_{lmk} \phi_{lmk}(\Omega,r) \Phi_{lmk} 
		+ \sum_{lm} \bval{\phi}_{lm}(\Omega,r) \bval{\Phi}_{lm}~,
\end{equation}
where the mode coefficients satisfy the reality condition
\begin{equation}
\label{appexp:reality}
	\Phi_{lmk}^\ast = (-1)^m \Phi_{l(-m)k}~, \qquad \bval{\Phi}_{lm}^\ast = (-1)^m \bval{\Phi}_{l(-m)}~.
\end{equation}
These follow from the properties of the spherical harmonics. Similarly, the generic expansion of vector fields reads 
\begin{align}
\label{appexp:vec.exp}
	\bA &= \sum_{lmk} \left[ \bv_{lmk}(\Omega,r) A^{V}_{lmk} + \bw_{lmk}(\Omega,r) A^{W}_{lmk} 
	+ \bx_{lmk}(\Omega,r) A^{X}_{lmk} \right] \\
\notag
	&\quad + \sum_{lm} \left[ \bval{\bv}_{lm} (\Omega,r) \bval{A}^{V}_{lm} 
	+ \bval{\bw}_{lm}(\Omega,r) \bval{A}^{W}_{lm}
	+ \bval{\bx}_{lm} (\Omega,r) \bval{A}^{X}_{lm} \right]~,
\end{align}
and the mode coefficients satisfy analogous reality conditions to \eqref{appexp:reality}. 
It turns out to be convenient to introduce the following combinations,
\begin{subequations}%
\label{appexp:A.hat.tilde}%
\begin{align}
	\widehat{A}_{lmk} &= \sqrt{\frac{l}{2l+1}} A^{V}_{lmk} - \sqrt{\frac{l+1}{2l+1}} A^{W}_{lmk}
		&&(l=1,2,3,\ldots),\\
	\widetilde{A}_{lmk} &= \sqrt{\frac{l+1}{2l+1}} A^{V}_{lmk} + \sqrt{\frac{l}{2l+1}}
		A^{W}_{lmk} && (l=0,1,2,\ldots).
\end{align}
\end{subequations}%
It is now possible to express the gradient of scalars, as well as the divergence and curl of vectors, in terms of the mode coefficients. Straightforward calculations yield 
\begin{subequations}
\label{appexp:grad}
\begin{gather}
\label{appexp:grad1}
	\widetilde{(\nabla \Phi)}_{lmk} = k \Phi_{lmk}~, \qquad
	\bval{(\nabla \Phi)}^W_{lm} = \frac{\sqrt{l(2l+1)}}{R} \bval{\Phi}_{lm}~,\\
\label{appexp:grad2}
	\widehat{(\nabla \Phi)}_{lmk} = (\nabla \Phi)^X_{lmk} 
	=\bval{(\nabla \Phi)}^V_{lm} = \bval{(\nabla \Phi)}^X_{lm}=0~,
\end{gather}
\end{subequations}
\begin{gather}
\label{appexp:div}
	(\nabla \cdot \bA)_{lmk} = -k \widetilde{A}_{lmk}~, \qquad
	\bval{(\nabla \cdot \bA)}_{lm} = -\frac{2l+3}R \sqrt{\frac{l+1}{2l+1}} \bval{A}^V_{lm}~,
\end{gather}
\begin{subequations}
\label{appexp:curl}
\begin{gather}
\label{appexp:curl1}
	\widehat{(\nabla \times \bA)}_{lmk} = k A^X_{lmk}~, \qquad
	(\nabla \times \bA)^X_{lmk} = k \widehat{A}_{lmk}~, \\
\label{appexp:curl2}
	\bval{(\nabla \times \bA)}^W_{lm}= - \frac{\sqrt{(l+1)(2l+1)}}{R} \bval{A}^X_{lm}~,\qquad
	\bval{(\nabla \times \bA)}^X_{lm}= \frac{2l+3}{R} \sqrt{\frac{l}{2l+1}} \bval{A}^V_{lm}~,\\
\label{appexp:curl3}
	\widetilde{(\nabla \times \bA)}_{lmk} =\bval{(\nabla \times \bA)}^V_{lm}=0~.
\end{gather}
\end{subequations}
It is an easy exercise to verify \eqref{appexp:helmholtz} using \eqref{appexp:grad}--\eqref{appexp:curl} and the identity $\nabla^2 \bA = \nabla (\nabla \cdot \bA) - \nabla \times (\nabla \times \bA)$.

Let us now choose a set $\{k\}$. If it were just for the scalar fields, this choice would be dictated by the boundary conditions at $r=R$, but the presence of the vector fields and the relations \eqref{appexp:grad}--\eqref{appexp:curl} make things somewhat more complicated. The simplest choice is 
\begin{equation}
\label{appexp:Bessel.bc}
	\{k\} = \left\{ k:j_l(kR)=0, k>0\right\}~.
\end{equation}
With \eqref{appexp:Bessel.bc} and the normalization constant
\begin{equation}
\label{appexp:clk}
	c_{lk} = \sqrt{\frac{R^3}2} j_{l+1}(kR)~,
\end{equation}
the functions \eqref{appexp:helmholtz.sols} are orthonormal with respect to the inner products
\begin{equation}
\label{appexp:norms}
	\left( \Phi,\Psi\right) = \int \rmd^2 \Omega \int\limits_0^R \rmd r\, r^2 \Phi^\ast \Psi~,
	\qquad
	\left( \bA,\bB \right) = \int \rmd^2 \Omega \int\limits_0^R \rmd r\, r^2 \bA^\ast\cdot \bB~.  
\end{equation}
The overlaps with the zero modes \eqref{appexp:helmholtz.zero.modes} are 
\begin{subequations}
\label{appexp:overlaps}
\begin{gather}
	\left( \phi_{lmk} ,\bval{\phi}_{l'm'} \right) = \left( \bx_{lmk} ,\bval{\bx}_{l'm'} \right) 
  	= \delta_{ll'} \delta_{mm'} \frac{\sqrt{2R}}{k}~,\\
    \left( \bv_{lmk} ,\bval{\bv}_{l'm'} \right) = \delta_{ll'} \delta_{mm'} \sqrt{\frac2R}\frac{2l+3}{k^2}~,\quad
    \left( \bw_{lmk} ,\bval{\bw}_{l'm'} \right) =0~.
\end{gather}
\end{subequations}
The last equation shows that the vector zero modes $\bval{\bw}_{lm}$ are needed for completeness. The other zero-modes, although not independent, are nevertheless useful for implementing boundary conditions that are not satisfied by the independent modes. Let us illustrate this for the scalar field. Consider a zero mode and expand it in terms of the complete basis,
\begin{equation}
\label{appexp:zero.mode.expand}
	\bval{\phi}_{lm} = \sum_{l'm'k} \left( \phi_{l'm'k} ,\bval{\phi}_{lm} \right) \phi_{l'm'k} 
	= \sum_k \frac{\sqrt{2R}}{k} \phi_{lmk}~.
\end{equation}
According to \eqref{appexp:grad}, its gradient is a vector field with coefficients $(\widetilde{\nabla\bval{\phi}_{lm}})_{lmk} = \sqrt{2R}$, which has infinite norm 
\begin{equation}
\label{appexp:inf.norm}
	2R\sum_k 1=\infty~,
\end{equation}
although $\left( \nabla \bval{\phi}_{lm}, \nabla \bval{\phi}_{lm} \right)$ is finite. The discrepancy is easily seen to arise from the boundary condition. Whereas the zero mode is non-zero at the boundary, the right hand side of \eqref{appexp:zero.mode.expand} vanishes there, but agrees with the zero-mode everywhere else, including at any inifinitesimal distance from the boundary. This implies that the $r$-derivative of the expansion diverges at the boundary, leading to \eqref{appexp:inf.norm}. The trick is to add a zero-norm state, 
\begin{equation}
\label{appexp:expansion.with.zero.mode}
	\Phi = \sum_{lmk} \Phi_{lmk} \phi_{lmk} + \sum_{lm} \bval{\Phi}_{lm} \left[ 
	\bval{\phi}_{lm} - \sum_k \frac{\sqrt{2R}}{k} \phi_{lmk} \right]~,
\end{equation}
and fix the coefficients $\bval{\Phi}_{lm}$ using the boundary values of $\Phi$. That the last term in \eqref{appexp:expansion.with.zero.mode} is a zero norm state follows from \cite{Sneddon:1960}
\begin{equation}
\label{appexp:Sneddon}
	\sum_{k} \frac1{k^2} = R^2 S_{2,l+1/2} = \frac{R^2}{2(2l+3)}~, \qquad
	\sum_{k} \frac1{k^4} = R^4 S_{4,l+1/2} = \frac{R^4}{2(2l+3)^2(2l+5)}~.
\end{equation}
This discussion justifies the use of the generic expansion \eqref{appexp:scalar.exp}, where $\Phi_{lmk}$ must now include the contribution from the zero-norm state. An analogous procedure applies for the vector zero modes $\bval{\bv}_{lm}$ and $\bval{\bx}_{lm}$.

We mention that the choice \eqref{appexp:Bessel.bc} for $\{k\}$ is not mandatory, but is the unique choice that makes all three sets of the bulk vector modes orthonormal. For example, we could have chosen
\begin{equation}
\label{appexp:Bessel.bc2}
	\{k\} = \left\{ k:j_{l-1}(kR)=0, k>0\right\}
\end{equation}
and
\begin{equation}
\label{appexp:clk2}
	c_{lk} = \sqrt{\frac{R^3}2} j_{l}(kR)~.
\end{equation}
This would have rendered the $\bw$-modes complete, but the $\bv$-modes would not form an orthogonal basis, although it would still be complete.

\end{appendix}

\section*{Acknowledgements}
This research was supported in part by the INFN research initiative ``STEFI''.


\begin{thebibliography}{10}

\bibitem{Dvali:2010jz}
G.~Dvali, G.~F. Giudice, C.~Gomez, and A.~Kehagias, ``{UV-Completion by
  Classicalization},'' \href{http://dx.doi.org/10.1007/JHEP08(2011)108}{{\em
  JHEP} {\bfseries 1108} (2011) 108},
\href{http://arxiv.org/abs/1010.1415}{{\ttfamily arXiv:1010.1415 [hep-ph]}}.

\bibitem{Dvali:2011aa}
G.~Dvali and C.~Gomez, ``{Black Hole's Quantum N-Portrait},''
  \href{http://dx.doi.org/10.1002/prop.201300001}{{\em Fortsch.Phys.}
  {\bfseries 61} (2013) 742--767},
\href{http://arxiv.org/abs/1112.3359}{{\ttfamily arXiv:1112.3359 [hep-th]}}.

\bibitem{Dvali:2012rt}
G.~Dvali and C.~Gomez, ``{Black Hole's 1/N Hair},''
  \href{http://dx.doi.org/10.1016/j.physletb.2013.01.020}{{\em Phys.Lett.}
  {\bfseries B719} (2013) 419--423},
\href{http://arxiv.org/abs/1203.6575}{{\ttfamily arXiv:1203.6575 [hep-th]}}.

\bibitem{Dvali:2012en}
G.~Dvali and C.~Gomez, ``{Black Holes as Critical Point of Quantum Phase
  Transition},'' \href{http://dx.doi.org/10.1140/epjc/s10052-014-2752-3}{{\em
  Eur. Phys. J.} {\bfseries C74} (2014) 2752},
\href{http://arxiv.org/abs/1207.4059}{{\ttfamily arXiv:1207.4059 [hep-th]}}.

\bibitem{Dvali:2013vxa}
G.~Dvali, D.~Flassig, C.~Gomez, A.~Pritzel, and N.~Wintergerst, ``{Scrambling
  in the Black Hole Portrait},''
  \href{http://dx.doi.org/10.1103/PhysRevD.88.124041}{{\em Phys. Rev.}
  {\bfseries D88} no.~12, (2013) 124041},
\href{http://arxiv.org/abs/1307.3458}{{\ttfamily arXiv:1307.3458 [hep-th]}}.

\bibitem{Dvali:2013eja}
G.~Dvali and C.~Gomez, ``{Quantum Compositeness of Gravity: Black Holes, AdS
  and Inflation},'' \href{http://dx.doi.org/10.1088/1475-7516/2014/01/023}{{\em
  JCAP} {\bfseries 1401} no.~01, (2014) 023},
\href{http://arxiv.org/abs/1312.4795}{{\ttfamily arXiv:1312.4795 [hep-th]}}.

\bibitem{Dvali:2015cwa}
G.~Dvali, ``{Quantum black holes},''
\href{http://dx.doi.org/10.1063/PT.3.2656}{{\em Phys. Today} {\bfseries 68}
  no.~1, (2015) 38--43}.

\bibitem{Dvali:2015aja}
G.~Dvali, ``{Non-Thermal Corrections to Hawking Radiation Versus the
  Information Paradox},''
\href{http://arxiv.org/abs/1509.04645}{{\ttfamily arXiv:1509.04645 [hep-th]}}.

\bibitem{Dvali:2015rea}
G.~Dvali, C.~Gomez, and D.~Lüst, ``{Classical Limit of Black Hole Quantum
  N-Portrait and BMS Symmetry},''
\href{http://arxiv.org/abs/1509.02114}{{\ttfamily arXiv:1509.02114 [hep-th]}}.

\bibitem{Binetruy:2012kx}
P.~Binetruy, ``{Vacuum energy, holography and a quantum portrait of the visible
  Universe},''
\href{http://arxiv.org/abs/1208.4645}{{\ttfamily arXiv:1208.4645 [gr-qc]}}.

\bibitem{Vikman:2012bx}
A.~Vikman, ``{Suppressing Quantum Fluctuations in Classicalization},''
  \href{http://dx.doi.org/10.1209/0295-5075/101/34001}{{\em Europhys.Lett.}
  {\bfseries 101} (2013) 34001},
\href{http://arxiv.org/abs/1208.3647}{{\ttfamily arXiv:1208.3647 [hep-th]}}.

\bibitem{Kovner:2012yi}
A.~Kovner and M.~Lublinsky, ``{Classicalization and Unitarity},''
  \href{http://dx.doi.org/10.1007/JHEP11(2012)030}{{\em JHEP} {\bfseries 1211}
  (2012) 030},
\href{http://arxiv.org/abs/1207.5037}{{\ttfamily arXiv:1207.5037 [hep-th]}}.

\bibitem{Alberte:2012is}
L.~Alberte and F.~Bezrukov, ``{Semiclassical Calculation of Multiparticle
  Scattering Cross Sections in Classicalizing Theories},''
  \href{http://dx.doi.org/10.1103/PhysRevD.86.105008}{{\em Phys.Rev.}
  {\bfseries D86} (2012) 105008},
\href{http://arxiv.org/abs/1206.5311}{{\ttfamily arXiv:1206.5311 [hep-th]}}.

\bibitem{Flassig:2012re}
D.~Flassig, A.~Pritzel, and N.~Wintergerst, ``{Black Holes and Quantumness on
  Macroscopic Scales},''
  \href{http://dx.doi.org/10.1103/PhysRevD.87.084007}{{\em Phys.Rev.}
  {\bfseries D87} (2013) 084007},
\href{http://arxiv.org/abs/1212.3344}{{\ttfamily arXiv:1212.3344}}.

\bibitem{Berkhahn:2013woa}
F.~Berkhahn, S.~Muller, F.~Niedermann, and R.~Schneider, ``{Microscopic Picture
  of Non-Relativistic Classicalons},''
  \href{http://dx.doi.org/10.1088/1475-7516/2013/08/028}{{\em JCAP} {\bfseries
  1308} (2013) 028},
\href{http://arxiv.org/abs/1302.6581}{{\ttfamily arXiv:1302.6581 [hep-th]}}.

\bibitem{Casadio:2013hja}
R.~Casadio and A.~Orlandi, ``{Quantum Harmonic Black Holes},''
  \href{http://dx.doi.org/10.1007/JHEP08(2013)025}{{\em JHEP} {\bfseries 1308}
  (2013) 025},
\href{http://arxiv.org/abs/1302.7138}{{\ttfamily arXiv:1302.7138 [hep-th]}}.

\bibitem{Muck:2014kea}
W.~Mück and G.~Pozzo, ``{Quantum portrait of a black hole with Pöschl-Teller
  potential},'' \href{http://dx.doi.org/10.1007/JHEP05(2014)128}{{\em JHEP}
  {\bfseries 05} (2014) 128},
\href{http://arxiv.org/abs/1403.1422}{{\ttfamily arXiv:1403.1422 [hep-th]}}.

\bibitem{Casadio:2014vja}
R.~Casadio, A.~Giugno, O.~Micu, and A.~Orlandi, ``{Black holes as
  self-sustained quantum states, and Hawking radiation},''
  \href{http://dx.doi.org/10.1103/PhysRevD.90.084040}{{\em Phys. Rev.}
  {\bfseries D90} no.~8, (2014) 084040},
\href{http://arxiv.org/abs/1405.4192}{{\ttfamily arXiv:1405.4192 [hep-th]}}.

\bibitem{Casadio:2015xva}
R.~Casadio, F.~Kuhnel, and A.~Orlandi, ``{Consistent Cosmic Microwave
  Background Spectra from Quantum Depletion},''
  \href{http://dx.doi.org/10.1088/1475-7516/2015/09/002}{{\em JCAP} {\bfseries
  1509} no.~09, (2015) 002},
\href{http://arxiv.org/abs/1502.04703}{{\ttfamily arXiv:1502.04703 [gr-qc]}}.

\bibitem{Casadio:2015bna}
R.~Casadio, A.~Giugno, and A.~Orlandi, ``{Thermal corpuscular black holes},''
  \href{http://dx.doi.org/10.1103/PhysRevD.91.124069}{{\em Phys. Rev.}
  {\bfseries D91} no.~12, (2015) 124069},
\href{http://arxiv.org/abs/1504.05356}{{\ttfamily arXiv:1504.05356 [gr-qc]}}.

\bibitem{Dvali:2014ila}
G.~Dvali, C.~Gomez, R.~S. Isermann, D.~Lüst, and S.~Stieberger, ``{Black hole
  formation and classicalization in ultra-Planckian $2\to N$ scattering},''
  \href{http://dx.doi.org/10.1016/j.nuclphysb.2015.02.004}{{\em Nucl. Phys.}
  {\bfseries B893} (2015) 187--235},
\href{http://arxiv.org/abs/1409.7405}{{\ttfamily arXiv:1409.7405 [hep-th]}}.

\bibitem{Dvali:2015jxa}
G.~Dvali, C.~Gomez, L.~Gruending, and T.~Rug, ``{Towards a Quantum Theory of
  Solitons},''
\href{http://arxiv.org/abs/1508.03074}{{\ttfamily arXiv:1508.03074 [hep-th]}}.

\bibitem{Barnich:2010bu}
G.~Barnich, ``{The Coulomb solution as a coherent state of unphysical
  photons},'' \href{http://dx.doi.org/10.1007/s10714-010-0984-6}{{\em
  Gen.Rel.Grav.} {\bfseries 43} (2011) 2527--2530},
\href{http://arxiv.org/abs/1001.1387}{{\ttfamily arXiv:1001.1387 [gr-qc]}}.

\bibitem{Mueck:2013mha}
W.~Mück, ``{On the number of soft quanta in classical field configurations},''
  \href{http://dx.doi.org/10.1139/cjp-2013-0712}{{\em Can. J. Phys.} {\bfseries
  92} no.~9, (2014) 973--975},
\href{http://arxiv.org/abs/1306.6245}{{\ttfamily arXiv:1306.6245 [hep-th]}}.

\bibitem{Mueck:2013wba}
W.~M{\"u}ck, ``{Counting Photons in Static Electric and Magnetic Fields},''
  \href{http://dx.doi.org/10.1140/epjc/s10052-013-2679-0}{{\em Eur.Phys.J.}
  {\bfseries C73} (2013) 2679},
\href{http://arxiv.org/abs/1310.6909}{{\ttfamily arXiv:1310.6909 [hep-th]}}.

\bibitem{Klauder:1996nx}
J.~R. Klauder, ``{Coherent state quantization of constraint systems},''
  \href{http://dx.doi.org/10.1006/aphy.1996.5647}{{\em Annals Phys.} {\bfseries
  254} (1997) 419--453},
\href{http://arxiv.org/abs/quant-ph/9604033}{{\ttfamily arXiv:quant-ph/9604033
  [quant-ph]}}.

\bibitem{Govaerts:1996zn}
J.~Govaerts, ``{Projection operator approach to constrained systems},''
  \href{http://dx.doi.org/10.1088/0305-4470/30/2/022}{{\em J. Phys.} {\bfseries
  A30} (1997) 603--617},
\href{http://arxiv.org/abs/hep-th/9606007}{{\ttfamily arXiv:hep-th/9606007
  [hep-th]}}.

\bibitem{Kapec:2015ena}
D.~Kapec, M.~Pate, and A.~Strominger, ``{New Symmetries of QED},''
\href{http://arxiv.org/abs/1506.02906}{{\ttfamily arXiv:1506.02906 [hep-th]}}.

\bibitem{He:2014cra}
T.~He, P.~Mitra, A.~P. Porfyriadis, and A.~Strominger, ``{New Symmetries of
  Massless QED},'' \href{http://dx.doi.org/10.1007/JHEP10(2014)112}{{\em JHEP}
  {\bfseries 10} (2014) 112},
\href{http://arxiv.org/abs/1407.3789}{{\ttfamily arXiv:1407.3789 [hep-th]}}.

\bibitem{Strominger:2014pwa}
A.~Strominger and A.~Zhiboedov, ``{Gravitational Memory, BMS Supertranslations
  and Soft Theorems},''
\href{http://arxiv.org/abs/1411.5745}{{\ttfamily arXiv:1411.5745 [hep-th]}}.

\bibitem{Watson}
G.~N. Watson, {\em {A Treatise on the Theory of Bessel Functions}}.
\newblock Cambridge University Press, 1966.

\bibitem{Hayashi:1981}
H.~Hayashi, ``Correction of the {Kneser-Sommerfeld} expansion formula,'' {\em
  J. Phys. Soc. Japan} {\bfseries 50} (1981) 1391. add.: \textbf{51} (1982)
  1324.

\bibitem{Vaccaro:1998}
V.~G. Vaccaro and L.~Verolino, ``Some remarks about the {Kneser-Sommerfeld}
  formula,'' {\em Il Nuovo Cimento} {\bfseries 113 B} (1998) 1527.

\bibitem{Kneser}
J.~C. C.~A. Kneser {\em Math. Ann.} {\bfseries LXIII} (1907) 447.

\bibitem{Sommerfeld}
A.~J.~W. Sommerfeld {\em Jahresber. Deutsch. Mat. Ver.} {\bfseries XXI} (1912)
  309.

\bibitem{Gradshteyn}
I.~S. Gradshteyn and I.~M. Ryzhik, {\em Table of Integrals, Series and
  Products}.
\newblock Academic Press, New York, 5~ed., 1994.

\bibitem{Barrera:1985}
R.~G. Barrera, G.~A. Estevez, and J.~Giraldo, ``Vector spherical harmonics and
  their application to magnetostatics,''
  \href{http://dx.doi.org/10.1088/0143-0807/6/4/014}{{\em Eur.J.Phys.}
  {\bfseries 6} (1985) 287}.

\bibitem{Hill:1954}
E.~L. Hill, ``The theory of vector spherical harmonics,''
  \href{http://dx.doi.org/http://dx.doi.org/10.1119/1.1933682}{{\em American
  Journal of Physics} {\bfseries 22} no.~4, (1954) 211--214}.

\bibitem{Abramowitz-Stegun}
M.~Abramowitz and I.~Stegun, {\em Handbook of Mathematical Functions}.
\newblock Dover, New York, fifth~ed., 1964.

\bibitem{Sneddon:1960}
I.~N. Sneddon, ``On some infinite series involving the zeros of bessel
  functions of the first kind,''
  \href{http://dx.doi.org/10.1017/S2040618500034067}{{\em Proceedings of the
  Glasgow Mathematical Association} {\bfseries 4} (1960) 144}.

\end{thebibliography}
\providecommand{\href}[2]{#2}\begingroup\raggedright\endgroup

\end{document}